\documentclass[aps,preprint,nofootinbib,preprintnumbers]{revtex4}

\usepackage{bm}
\usepackage{graphicx}    % needed for including graphics e.g. EPS, PS
\usepackage{amsfonts}
\usepackage{amsmath}
\usepackage[dvips]{color}
\usepackage[normalem]{ulem}
\usepackage{amsmath}
\usepackage{enumerate}
\usepackage{amsfonts}
\usepackage{epsfig}

\newcommand{\be}{\begin{equation}}
\newcommand{\ee}{\end{equation}}
\newcommand{\bea}{\begin{eqnarray}}
\newcommand{\eea}{\end{eqnarray}}

\newcommand{\bln}{\begin{align}}
\newcommand{\eln}{\end{align}}
\newcommand{\bst}{\begin{split}}
\newcommand{\est}{\end{split}}
\newcommand{\bi}{\begin{itemize}}
\newcommand{\ei}{\end{itemize}}
\newcommand{\ben}{\begin{enumerate}}
\newcommand{\een}{\end{enumerate}}

\def\ov{\over}
\def\le{\left}
\def\ri{\right}
\def\ha{{1\over 2}}

\def\al{{\alpha}}

\def\vev#1{\langle#1\rangle}

\def\tr{{\rm tr}}

\def\ep{{\epsilon}}

\newcommand{\p}{\partial}

\newcommand\ga{{\ensuremath{{\gamma}}}}
\newcommand\Ga{{\ensuremath{{\Gamma}}}}

\newcommand\lam{\lambda}

\newcommand\om{\omega}

\newcommand\De{{\ensuremath{{\Delta}}}}

\def\lam{{\lambda}}

\def\eeq{\end{equation}}

\newcommand\sD{{\ensuremath{{\mathcal D}}}}

\newcommand\sM{{\ensuremath{{\mathcal M}}}}
\newcommand\sN{{\ensuremath{{\mathcal N}}}}
\newcommand\sO{{\ensuremath{{\mathcal O}}}}

\newcommand\sJ{{\mathcal J}}
\newcommand\sS{{\mathcal S}}

\newcommand\sT{{\mathcal T}}

\newcommand\bpsi{{\bar \psi}}

\newcommand\da{{\dagger}}

\begin{document}

\title {Real-time response in AdS/CFT with application to spinors\footnote{To appear
in the Proceedings of the 4-th RTN ``Forces-Universe'' Workshop.}}

\preprint{MIT-CTP 4022}

\author{Nabil Iqbal and Hong Liu}

\affiliation{Center for Theoretical Physics, \\
Massachusetts
Institute of Technology, \\
Cambridge, MA 02139
}

%\date{May, 2008}

\bigskip
\bigskip

\vspace{5cm}

\begin{abstract}

We discuss a simple derivation of the real-time AdS/CFT prescription
as an analytic continuation of the corresponding problem in Euclidean signature. We then extend the formalism to spinor operators and apply it to the examples of real-time fermionic correlators in CFTs dual to pure AdS and the BTZ black hole.

\end{abstract}

\maketitle
%\newpage

%\title{A note on real-time correlators from AdS/CFT}
%\author{Nabil Iqbal and Hong Liu}
%\maketitle
%\abstract{We review a simple derivation of the real-time AdS/CFT prescription as an analytic continuation of the corresponding problem in Euclidean signature. We then extend the formalism to spinor operators and apply it to the demonstrative examples of real-time fermionic correlators in CFTs dual to pure AdS and the BTZ black hole. }

\section{Introduction}

The AdS/CFT correspondence~\cite{Maldacena:1997re,Gubser:1998bc,Witten:1998zw}
has provided important tools for studying strongly coupled systems
by relating them to classical gravity (or string) systems. For understanding properties of such a system at finite temperature or finite density, real-time two-point correlation functions of composite operators are of great importance, as they encode information about collective behavior of the system such as transport, (the possible existence of) quasi-particles, etc.

A simple prescription for calculating retarded two-point functions in AdS/CFT was first proposed by Son and Starinets some years ago in~\cite{Son:2002sd}.
This prescription, which was justified in different ways in the literature (see e.g.~\cite{Herzog:2002pc,Marolf:2004fy,gubser:2008,skenderis,van_rees:2009}), has been instrumental in extracting many important insights into strongly interacting many-body systems from AdS/CFT.

In~\cite{iqbal:2008} we observed that the prescription of~\cite{Son:2002sd} could be reformulated in terms of boundary values of the canonical momenta of bulk fields by treating the AdS radial direction as ``time''. This reformulation has both conceptual and practical advantages; it expresses real-time response in terms of objects with intrinsic physical and geometric meaning. This is, for example, essential for the proof in \cite{iqbal:2008} that various boundary transport coefficients can be solely expressed in terms of quantities evaluated at the horizon.

In this note we further shed light on the formulation of \cite{iqbal:2008} by showing that it follows simply from an analytic continuation of the original Euclidean formulation of AdS/CFT. At the end of our investigation, we came to realize that a similar procedure has in fact been used earlier~\cite{gubser:2008} to justify the prescription of~\cite{Son:2002sd}. Nevertheless, we feel that our presentation may still be of some value, as it highlights the importance of the canonical momenta as the response of the dual operator even in real-time.

We then show that our formalism may be extended to spinors, which involve interesting subtleties as we are now dealing with a first order system. We work out an explicit prescription for calculating real-time retarded correlators of fermionic operators, which is important, e.g., for probing quasi-particle structure associated with Fermi surfaces from gravity~\cite{Lee:2008xf,vegh}.

The plan of the paper is as follows. In section~\ref{sec:cont} we state and derive our reformulation of the real-time prescription for retarded Green's functions in AdS/CFT.
In section~\ref{sec:fer} we work out in detail the application of this prescription to spinor fields in an asymptotic AdS spacetime. In section~\ref{sec:ex} we further illustrate the prescription using two exactly solvable examples: pure AdS and BTZ black hole.
We conclude in section~\ref{sec:con} with a brief summary. In Appendix~\ref{app:A}, we state our conventions for various Euclidean and retarded Green's functions, while Appendices~\ref{app:M} and~\ref{app:B} contain some technical details related to the main text.

\section{Derivation of a real-time prescription for retarded correlators in AdS/CFT} \label{sec:cont}

In this section we first state and then derive our reformulation of the real-time prescription for retarded Green's functions in AdS/CFT.
For illustration, we consider a scalar operator $\sO$ which is dual to a massless bulk scalar field $\phi$.  Massive modes contain extra divergences; while harmless, these complicate the discussion and are discussed in Appendix~\ref{app:M}. The generalization to tensors is self-evident, and the generalization to spinors is discussed in detail in the next section.

To be specific, we will consider a scalar action
\be \label{lorc}
S = -\ha \int d^{d+1} x \, \sqrt{-g} \, (\p \phi)^2 + \cdots
\ee
on a background spacetime metric of the form
\be \label{gmetr}
ds^2 = - g_{tt} dt^2 + g_{rr} dr^2 + g_{ii} d\vec x^2
\equiv g_{rr} dr^2 + g_{\mu \nu} dx^\mu dx^\nu  \ .
\ee
The boundary is taken at $r=\infty$, where various components of the metric have the asymptotic behavior of AdS with unit radius,
\be \label{lagr}
g_{tt}, g^{rr},  g_{ii} \approx {r^2}, \qquad r\to \infty \ .
\ee
We will assume that the above metric has a horizon of some sort (to be more explicit below) in the interior.\footnote{In a spacetime which does not have a horizon (e.g. global AdS), i.e. completely regular in the interior, $\phi$ has a discrete spectrum. The corresponding boundary theory retarded function for $\sO$ can then be written as a discrete sum of delta functions and thus does not have an interesting analytic structure.}

We also assume that the theory is  translationally invariant in $x^\mu$ directions (i.e, all metric components depend on $r$ only), and
work in momentum space along these directions, e.g.
\be \label{rey}
\phi (r, x^\mu) = \phi (r, k_\mu) \, e^{-i \om t + i \vec k \cdot \vec x },
\qquad k_\mu = (-\om, \vec k) \ .
\ee
We will frequently need to analytically continue to Euclidean signature via
\be \label{EucA}
t \to - i \tau, \qquad  \om \to i \om_E , \qquad i S \to - S_E
\ee
where the Euclidean action $S_E$ is given by
\be \label{eucA}
S_E = \ha \int d^{d+1} x \, \sqrt{g} \, (\p \phi)^2 + \cdots \ .
\ee

\subsection{Prescription}

We begin by simply stating the prescription of~\cite{iqbal:2008} for computing the retarded correlator $G_R (k_\mu)$ for $\sO$. In momentum space this reads:\footnote{The reason we have~\eqref{eqGr} instead of $ G_R(k_\mu) = \left(\lim_{r\to\infty} \frac{\delta \Pi|_{ \phi_R}}{\delta \phi_0} \right)\biggr|_{\phi_0 = 0}$ is due to subtleties in taking the limit $r \to \infty$. An example is the massive scalar case discussed in Appendix~\ref{app:M}. Also, we found in \cite{iqbal:2008} that the fact that all quantities in \eqref{eqGr} remain meaningful in the bulk at arbitrary radius is helpful for a physical interpretation of this formula.}
\be
G_R(k_\mu) = \left(\lim_{r\to\infty}\frac{\Pi(r;k_\mu)|_{ \phi_R}}{\phi_R (r,k_\mu)}\right)\biggr|_{\phi_0 = 0}
\label{eqGr}
\ee
where $\Pi$ is the canonical momentum conjugate to $\phi$ with respect to a foliation in the $r$-direction.
$\phi_R (r, k_\mu)$ is the solution to the equations of motion for $\phi$ which
is in-falling at the horizon and satisfies the boundary condition $\lim_{r \to \infty}
\phi_R (r, k_\mu) \to \phi_0 (k_\mu) $. The notation $\Pi(r;k_\mu)|_{ \phi_R}$ in equation~\eqref{eqGr} indicates that $\Pi$ should be evaluated
on the solution $\phi_R$. Finally, the subscript $\phi_0 = 0$ means that in evaluating this ratio one should only take the part that is independent of $\phi_0$; in an interacting bulk theory $\Pi$ and $\phi_R$ will typically contain higher powers of $\phi_0$ which are not relevant for two-point functions.

Note that~\eqref{eqGr} can also be equivalently (in fact more generally) written as
\be \label{oneP}
\vev{\sO (k_\mu)}_{\phi_0} = \lim_{r \to \infty} \Pi(r;k_\mu)|_{ \phi_R}
\ee
where $\vev{\sO (k_\mu)}_{\phi_0}$ denotes the response of the system to external perturbations generated by adding $\int d^d x \, \phi_0 (x) \sO (x)$ to the boundary theory action. Equation~\eqref{eqGr} is the linearized limit of~\eqref{oneP}.\footnote{To see this more explicitly, it is helpful to recall the standard result from linear response theory that if one considers a system in equilibrium at $t \to -\infty$ and then perturbs its action with the term $\int d^d x \, \phi_0 (x) \sO (x)$, the one-point function of $\sO$ in the presence of the source is given (to first order in $\phi_0$) by
$$
\langle \sO(k_{\mu}) \rangle_{\mathrm{\phi_0}} = G_R(k_{\mu})\phi_0(k_{\mu}),
$$
where $G_R$ is the retarded correlator of $\sO$.
}

An alert reader will immediately recognize that~\eqref{oneP} is the exact Lorentzian analog of the standard prescription for boundary correlation functions in Euclidean signature as formulated in~\cite{Gubser:1998bc,Witten:1998zw}. Indeed we will now show that ~\eqref{eqGr} and~\eqref{oneP} (at the linearized level) can be derived from the original prescription of~\cite{Gubser:1998bc,Witten:1998zw} using a simple analytic continuation, even though they do not appear to follow directly from an action principle themselves.

\subsection{Analytic continuation}

First recall that a retarded correlator is analytic in the upper half complex-$\om$ plane, and that the value of $G_R (\om, \vec k)$ along the upper imaginary $\om$-axis gives us the Euclidean correlator $G_E$, i.e.\footnote{$G_E (\om_E)$ for $\om_E < 0$ is obtained from the advanced function. See Appendix~\ref{app:A} for our conventions on the definitions of Euclidean and retarded functions.},
\be \label{ana1}
G_E (\om_E, \vec k) =
G_R (i \om_E, \vec k) \qquad \om_E > 0 \ .
\ee
This expression applies both for finite and zero temperature, and both for bosonic and fermionic operators, with $\om_E$ only taking discrete values at finite temperature. Equation~\eqref{ana1} can now be inverted to obtain
$G_R (\om, \vec k)$ as
\be
G_R(\om, \vec k) = G_E(\om_E , \vec k) \biggr|_{\om_E= -i (\om + i \epsilon)} \ .
%\qquad G_A(k_0) = -G_E(\omega = ik_0 + \epsilon)
\label{contGR}
\ee
Note that at finite temperature $\om_E$ only takes discrete values, and so the analytic continuation can be tricky. We will simply ignore this issue below, assuming that $G_R$ is sufficiently well-behaved that such an analytic continuation is possible.

We now look at the central prescription of AdS/CFT in Euclidean signature~\cite{Gubser:1998bc,Witten:1998zw} (for both
zero and finite temperature)
\be \label{cp}
\left\langle \exp\left[\int d^dx\;\phi_0 (x) \sO (x) \right]\right\rangle_\mathrm{QFT} = e^{-S_{\mathrm{grav}}[\phi_0]}
\ee
where $S_{\mathrm{grav}}[\phi_0]$ is the bulk action for $\phi$ (which may include necessary boundary terms) evaluated at the classical Euclidean solution $\phi_E$ which is regular in the interior and satisfies boundary condition $\lim_{r \to \infty}
\phi_E (r; x) \to \phi_0 (x)$. From~\eqref{cp}, one finds that one point function of $\sO$ in the presence of the source $\phi_0$ can be written as
\be
\label{defexpO}
\langle \sO(x) \rangle_{\phi_0} = -{\delta S_{\rm grav} \ov \delta \phi_0 (x)} = - \lim_{r\to\infty}\Pi_E (r,x)|_{\phi_E}
\ee
and its Fourier transform
\be \label{four1}
\langle \sO(\om_E, \vec k) \rangle_{\phi_0}  = -\lim_{r\to\infty}\Pi_E (r,\om_E, \vec k)|_{ \phi_E} \ .
\ee
In~\eqref{defexpO}, $\Pi_E $ is the canonical momentum for $\phi$ in Euclidean signature and should be evaluated at the classical solution $\phi_E$. The
second equality of~\eqref{defexpO} follows from the well known fact in classical mechanics that the derivative of an on-shell action with respect to the boundary value of a field is simply equal to the canonical momentum conjugate to the field, evaluated at the boundary. Note that in general the limit $r \to \infty$ in~\eqref{four1} is non-trivial and requires careful renormalization, which was developed systematically in~\cite{skenderis:2004a,skenderis:2004b}. Similar cautionary remarks apply to other formulas below (in both Lorentzian and Euclidean signature).

At linear level in $\phi_0$, one then finds that
\be \label{eru}
G_E(\om_E, \vec{k}) = -\left(\lim_{r\to\infty}\frac{\Pi_E (r,\om_E, \vec k)|_{ \phi_E}}{\phi_E (r,\om_E,\vec{k})}\right)
\biggr|_{\phi_0 =0}
\ee
In Euclidean signature, the above procedure is unambiguous as $\phi_E$ is unique
(at least for sufficiently small $\phi_0$); there will be only one solution that both satisfies the boundary conditions at infinity and is regular in the interior. This is not the case in Lorentzian signature, where in the interior one typically finds two oscillatory solutions, both of which are regular.

Once $G_E (\om_E, \vec k)$ is obtained, one can of course obtain $G_R$ using~\eqref{contGR}. Here, we would like to argue that one can in fact obtain an intrinsic prescription for retarded functions by analytically continuing $\phi_E (r, \om_E, \vec k)$ directly in the bulk to Lorentzian signature\footnote{A similar analytic continuation has been used earlier in~\cite{gubser:2008} to justify the prescription of~\cite{Son:2002sd}.}. More explicitly,
introducing
%one can show that $\phi_R$ introduced below~\eqref{eqGr} is precisely related to $\phi_E$ by analytic continuation
\be \label{dfR}
\phi_R (\om,\vec k) = \phi_E (\om_E, \vec k)\biggr|_{\om_E= -i (\om + i \epsilon)}
\ee
and then from~\eqref{contGR} and \eqref{eru}, we obtain simply\footnote{Note the relative minus sign between the first and second equality below is due to that in Euclidean signature $\Pi_E = \sqrt{g} g^{rr} \p_r \phi$ (see~\eqref{eucA}) and in Lorentzian signature $\Pi = - \sqrt{-g} g^{rr} \p_r \phi$ (see~\eqref{lorc}) with an extra minus sign.}
\be
G_R(\omega, k_\mu) = - \lim_{r\to\infty}\frac{\Pi_E (r,\om_E, \vec k)|_{ \phi_E}}{\phi_E (r,\om_E,\vec{k})}\biggr|_{\om_E= -i (\om + i \epsilon)} = \lim_{r\to\infty}\frac{\Pi(r;\omega, \vec{k})|_{ \phi_R}}{\phi_R (r,\omega, \vec{k})},
\ee
which gives \eqref{eqGr}. Similarly, \eqref{four1} becomes~\eqref{oneP} under the same analytic continuation. To complete the derivation we still need to show that $\phi_R$ obtained in~\eqref{dfR} satisfies the boundary conditions at the horizon and infinity as stated below~\eqref{eqGr}. Since analytic continuation does not change the boundary conditions at infinity, we need only show that the function on the right-hand side of~\eqref{dfR} satisfies the in-falling condition at the horizon. We do this explicitly by examining the behavior of $\phi_E$ near different types of horizons\footnote{Note that the following discussion applies to any mass, as the mass term is never important at the horizon with a nonvanishing $\om$.}:

\bi

\item {\it Non-degenerate horizon}: consider a Euclidean black hole background of the form
\be \label{bhr}
ds^2 = f(r) d\tau^2 + \frac{1}{f(r)} dr^2 + a(r)^2 d\vec{x}^2
\ee
where $f(r)$ has a simple zero at $r = r_0$, i.e. $f(r) \sim \frac{4\pi}{\beta} (r-r_0)$
and $\tau$ is periodic in $\beta$. From the wave equation of $\phi$, one finds the following near-horizon solutions
\be
\phi_+(r) \sim (r-r_0)^{+\frac{\om_E \beta}{4\pi}} \qquad \mbox{or} \qquad \phi_-(r) \sim (r-r_0)^{-\frac{\om_E \beta}{4\pi}}
\label{soln1} \ .
\ee
When the real part of $\om_E > 0$ (which is always the case when evaluating a retarded correlator by \eqref{contGR}), $\phi_E \sim \phi_+$ and thus we find that
\be
\phi_E (\om_E, \vec k)\biggr|_{\om_E= -i (\om + i \epsilon)}  \sim (r-r_0)^{-\frac{i\om \beta}{4\pi}}
\ee
which is precisely the behavior of an in-falling wave, since from~\eqref{rey} it
leads in coordinate space to a wave of the form $e^{-i \om t} (r-r_0)^{-\frac{i\om \beta}{4\pi}}$.

\item {\it Degenerate horizon:} For a degenerate horizon (e.g. of an extremal black hole), we have a metric of the form~\eqref{bhr} except that $f$ now has a double pole at $r=r_0$, i.e. $f \sim c(r-r_0)^2$, where $c > 0$. The discussion is similar to that above except that the near-horizon solution can now be shown to take the form
\be
\phi_E (\om_E, \vec k)\biggr|_{\om_E= -i (\om + i \epsilon)}  \sim e^{i \om \ov  c(r-r_0)},
\ee
again corresponding to an in-falling wave.

\item {\it Poincar\'e horizon}: finally let us consider a  Poincar\'e horizon of the form (near $z \to \infty$)
\be
ds^2 = \frac{1}{z^2}\left[d\tau^2 + d\vec{x}^2 + dz^2\right]
\ee
Then the near horizon solutions for $\phi$ are given by
\be
\phi_\pm \sim e^{\pm k z}, \qquad k = \sqrt{\om_E^2 + \vec k^2}, \qquad z \to + \infty
\ee
Note that if we take the branch of the square root that is positive, then the regular solution $\phi_E \sim \phi_-$. We now perform the analytic continuation  $\om_E = -i (\om + i \ep)$, under which the branch of the square root with positive real part becomes
\be
\label{pep}
k = \begin{cases} -i \sqrt{\om^2 - \vec k^2} & \om > |\vec k| \cr
                 i \sqrt{\om^2 - \vec k^2} & \om < -  |\vec k|
                 \end{cases}
\ee
where we have assumed a timelike momentum $|\om| > |\vec{k}|$ (as otherwise the problem is the same as in the Euclidean case). When plugged into $\phi_- \sim e^{-kz}$, Equation~\eqref{pep} again describes an in-falling wave for both signs of $\om$.

\ei
The key point here is that the $+i\epsilon$ in the analytic continuation \eqref{contGR} is important; it guarantees that the solution that was regular at the horizon in Euclidean signature becomes the solution that is in-falling at the horizon in Lorentzian signature.

Note that while the Euclidean prescription~\eqref{cp}--\eqref{eru} for computing correlation functions follow from an action principle, this is not the case for the Lorentzian prescription~\eqref{eqGr} and~\eqref{oneP}, which cannot be obtained by taking functional derivatives
of the on-shell Lorentzian action. While analytic continuation of a Euclidean solution yields a Lorentzian solution as in \eqref{dfR}, analytic continuation of a Euclidean on-shell \emph{action}--which necessarily involves an integral over the full spacetime manifold--does not necessarily yield the correct Lorentzian action. The Lorentzian manifold generically differs from the Euclidean one in many crucial aspects such as the topology, the number of boundaries, etc., making the implementation and interpretation of such a procedure much more subtle. On the other hand, the analytic continuation of the canonical momentum as in our prescription requires only a \emph{local} analytic continuation on the boundary of the manifold and is thus much simpler.

\subsection{Vector and Tensor Operators}

Before concluding our discussion of bosonic operators, we point out that in this formalism the tensor structure of the real-time response follows naturally without needing to decompose the system mode by mode into a series of scalar wave equations. Let us imagine first a bulk vector field $A_M$ with Maxwell action and gauge coupling $g_{\mathrm{eff}}$; this is dual to a conserved current $\sJ^{\mu}$, and the expression \eqref{oneP} then becomes
\be
\langle \sJ^{\mu} \rangle = -\lim_{r\to\infty}\frac{1}{g_{\mathrm{eff}}^2}\sqrt{-g}F^{r \mu}
\ee
where the expression on the right-hand side is the momentum conjugate to $A_\mu$. Similarly, if we consider the field theory stress tensor $\sT^{\mu\nu}$, we find that
\be \label{stens}
\langle \sT^{\mu \nu} \rangle =  \lim_{r\to\infty} {\sqrt{-\gamma} \ov 16 \pi G_N} \le(K^{\mu \nu} - \gamma^{\mu \nu}
K^\lam_\lam \ri),
\ee
correctly captures the real-time response of the system to metric perturbations. Here $\gamma_{\mu\nu}$ and $K_{\mu\nu}$ are the induced metric and extrinsic curvature on each constant-r slice, and the expression on the right-hand side is the Brown-York stress tensor \cite{brownyork, bk}, i.e. precisely the momentum conjugate to a gravitational perturbation $h_{\mu\nu}$. The validity of these formulas even in real-time was a key element in the results relating AdS/CFT to the membrane paradigm in \cite{iqbal:2008}.

\section{Retarded correlators for fermionic operators}
\label{sec:fer}

In this section we generalize the prescription~\eqref{eqGr} and~\eqref{oneP} to fermionic operators. We first consider the calculation of Euclidean functions in an asymptotic AdS geometry, reviewing (and slightly generalizing) the earlier results of~\cite{Henningson:1998cd,Mueck:1998iz,henneaux:1999}.
We then discuss the prescription for real-time retarded correlators. In the next section we use the prescription to calculate retarded correlation functions for fermonic operators in field theories dual to the pure AdS and BTZ geometries, where closed-form expressions can be obtained.

We consider a boundary theory fermionic operator $\sO$ which is dual to a
spinor field $\psi$ in the bulk\footnote{Note that in this section $\sO$ is fermionic whereas in Section \ref{sec:cont} $\sO$ was bosonic; we apologize for the abuse of notation.}. We will suppress all spinor indices. Since we are interested in two point functions of $\sO$, it is enough to consider the quadratic part of the action for $\psi$, which can be written as
\be
S =  \sN \int d^{d+1} x \sqrt{-g} \, i (\bar \psi \Ga^M D_M \psi  - m \bpsi \psi) + S_{{\rm bd}} \label{diracac}
\ee
where $\bpsi = \psi^\da \Ga^t$ and
\be \label{nek}
D_M  = \p_M + {1 \ov 4} \om_{ab M} \Ga^{ab} \ .
\ee
$S_{\rm bd}$ denotes the boundary terms required to ensure that the total action has a well defined variational principle~\cite{henneaux:1999} and is briefly discussed in Appendix~\ref{app:B}. In~\eqref{diracac}--\eqref{nek}  $M , N \cdots$ denote
abstract spacetime indices and $a,b, \cdots$ denote
abstract tangent space indices. Gamma matrices with a specific index (like $\Ga^{t}, \Ga^r, \Ga^i$ as opposed to those with an abstract index $\Ga^M$) always correspond to those in the tangent frame. The boundary theory gamma matrices will be denoted by $\ga^\mu$. 
The background metric is taken to be~\eqref{gmetr} with the associated asymptotics. We will be working in momentum space with the Fourier transform of $\psi$ denoted by $\psi (r, k_\mu)$. Finally, the action is normalized by a factor $\sN$; we will ignore this factor in the following, as it simply contributes an overall factor in front of all boundary theory correlators, but one can show that its sign is fixed by bulk unitarity.\footnote{To see this, consider canonically quantizing the action \eqref{diracac} by imposing the equal-time anticommutation relation $\{\psi(x), \pi(x')\} = i\delta^{(d)}(x,x')$, where here $\pi$ is the momentum conjugate to $\psi$ with respect to \emph{time}. We find that $\pi = - i\sN\sqrt{-g}\psi^{\dagger}$. If we now require that the anticommutator $\{\psi, \psi^{\dagger}\}$ be \emph{positive}--a requirement in a Hilbert space with states of positive norm--then we find that this fixes the sign of $\sN$ to be \emph{negative}. Reinstating this factor of $\sN$ in front of retarded correlators found in this paper results in positive spectral densities in the boundary theory, as required by boundary unitarity.}

The analytic continuation to Euclidean signature is as in~\eqref{EucA} with also
\be
\ga^t \to - i \ga^\tau, \qquad \Ga^{t} \to - i \Ga^{\tau} ,  \qquad
\bar \psi \to -i \bar \psi
\ee
and thus the Euclidean action following from \eqref{diracac}
\be \label{diracacE}
S_E = - \int d^{d+1} x \sqrt{g} \,  (\bar \psi \Ga^M \sD_M \psi  - m \bpsi \psi)
+ S_{\rm bd} \ .
\ee

\subsection{Euclidean correlators}

We first review how the Euclidean prescription~\eqref{cp}--\eqref{eru} works for
a spinor operator, where we now have
\be \label{vne}
\left\langle \exp\left[\int d^dx\; \le(\bar \chi_0 \sO + \bar O \chi_0 \ri)  \right]\right\rangle_\mathrm{QFT} = e^{- S_{\rm grav} [\chi_0, \bar \chi_0]} .
\ee
To find the right hand side of~\eqref{vne}, we need to construct a $\psi_E$ which is regular in the interior and satisfies the boundary condition $\lim_{r \to \infty} \psi_E = \chi_0$. Attempting to interpret this equality, we realize that the situation here is a bit more subtle than for a scalar field: $\psi$ and $\chi_0$ are spinors of different spacetime dimensions, and thus may have a different number of components. Also, the action for $\psi$ contains only one derivative and imposing Dirichlet boundary conditions for such a first order system requires more care. It turns out that these two issues are intimately related.

To see this, it is convenient to
decompose $\psi$ in terms of eigenvalues of $\Ga^r$, with
\be \label{deC}
\psi = \psi_+ + \psi_-, \qquad \psi_\pm = \Ga_\pm \psi, \qquad \Ga_\pm = {1 \ov 2} \le(1 \pm \Ga^r \ri) \
\ee
where from \eqref{diracacE} we find the corresponding Euclidean canonical momentum\footnote{For fermions we use only the Euclidean canonical momenta, and have thus omitted the subscript $E$}
$\Pi_{\pm}$ (in $r$-slicing) conjugate to $\psi_\pm$ to be
\be \label{deffermpi}
\Pi_{+}   = -  \sqrt{g} g_{rr}^{-\ha}  \bpsi_- , \qquad \Pi_{-}   =    \sqrt{g} g_{rr}^{-\ha}  \bpsi_+ \ .
\ee
Thus we see that $\psi_\pm$ are conjugate to each other. If we supplied Dirichlet boundary conditions for both of them at infinity, then it would completely fix the solution everywhere; however this is obviously incorrect, as (generically) this solution would not be regular in the interior. Instead we should \emph{begin} by demanding regularity in the interior; this then leaves us with the freedom to impose boundary conditions $\chi_0$ for either $\psi_+$ or $\psi_-$, i.e. only for half of the components of $\psi$.

When the boundary theory dimension $d$ is even, we can choose the bulk Gamma matrices
\be \label{evch}
\Ga^\mu =  \ga^\mu, \qquad \Ga^r = \ga^{d+1}
\ee
where $\ga^{d+1}$ is the analogue of $\ga^5$ for $d=4$. Thus from the $d$-dimensional point of view, the two components $\psi_{\pm}$ transform like $d$-dimensional Weyl spinors of opposite chirality. As the boundary value of one of $\psi_{\pm}$, $\chi_0$ (and so also $\sO$) is then a $d$-dimensional boundary spinor of definite chirality. Thus for $d$ even a Dirac spinor $\psi$ in the bulk is mapped to a chiral spinor $\sO$ on the boundary.

When $d$ is odd, it is convenient to work with the representation
\be \label{chga}
\Gamma^r = \left(\begin{tabular}{cc}$1$ & $0$ \\ $0$ & $-1$\end{tabular}\right) \qquad \Gamma^{\mu} = \left(\begin{tabular}{cc}0 & $\gamma^{\mu}$ \\ $\gamma^{\mu}$ & 0\end{tabular}\right),
\ee
which can be easily seen to satisfy the $d+1$ dimensional Clifford algebra. In this basis we can see that the two components $\psi_{\pm}$ each transform as a $d$-dimensional Dirac spinor; thus for $d$ odd $\chi_0$ and $\sO$ are both Dirac spinors. In all dimensions the number of components of $\sO$ is always half of that for $\psi$.

To decide on which of $\psi_\pm$ to impose Dirichlet boundary conditions, we must
examine their asymptotic behavior at large $r$, which can be worked out by solving the Dirac equation for $\psi$ in the region $r \to \infty$ (with metric given by~\eqref{lagr}). This computation is performed in more detail in Section \ref{sec:pureads} and here we discuss only the asymptotic behavior of $\psi$, which  is given by
\be \label{r0}
\psi_+ (r, k)= A (k)  r^{-{d \ov 2} + m} + B(k)  r^{-{d \ov 2} - m -1},
\qquad
\psi_- = C(k) r^{-{d \ov 2} + m -1} + D (k)r^{-{d \ov 2} - m }, \quad r \to \infty
\ee
Plugging this expansion back into the Dirac equation we find the following relations between the expansion coefficients
\be \label{r1}
D = - {i \ga \cdot k \ov k^2} (2m +1) B, \qquad
C =  {i \ga \cdot k \ov  2m -1} A, \qquad \ga \cdot k = \ga^\mu k_\mu, \quad k^2 = k_\mu k^\mu \ .
\ee
Using \eqref{deffermpi}, we see that the corresponding canonical momenta behave as
\be \label{r3}
\Pi_{+} = -\bar C r^{{d \ov 2} + m -1} - \bar D  r^{{d \ov 2} - m }, \qquad
\Pi_{-} = \bar A r^{{d \ov 2} + m } + \bar B  r^{{d \ov 2} - m -1} , \qquad r \to \infty \ .
\ee
Note that in deriving~\eqref{r1} we have used~\eqref{evch} or~\eqref{chga} for $d$ even or odd.

Notice that as we take $m \to -m$, we simply exchange the role of $\psi_\pm$, with $A \leftrightarrow D$ and $B \leftrightarrow C$. We can thus restrict our attention to $m \geq 0$.
Among all terms in~\eqref{r0}, the term with coefficient $A$ is dominant. We thus should the impose boundary conditions
\be \label{rio}
A = \chi_0, \quad {\rm i.e.} \quad \lim_{r \to \infty} r^{{d \ov 2} -m} \psi_+ = \chi_0 \ .
\ee
Now taking a derivative with respect to $\chi_0$ on both sides of~\eqref{vne}, one finds that as in~\eqref{defexpO}, the response of $\bar{\sO}$ is formally given by the conjugate momentum $\Pi_{+}$.\footnote{See Appendix~\ref{app:B} for an explicit discussion of the variation of $S_{\rm grav}$.} However, for general $m$, as in the case of a massive scalar discussed in Appendix~\ref{app:M}, one should define the boundary limit carefully. In view of the second equation in~\eqref{rio}, we thus find
\be \label{ebe}
\vev{\bar \sO}_{\chi_0} =- \lim_{r \to \infty} r^{m - {d \ov 2}} \Pi_{+} , \quad {\rm i.e.} \quad \vev{\bar \sO}_{\chi_0} = \bar D \ .
\ee
Note that in obtaining the second equation above we have extracted only the finite terms in the right hand side of the first equation\footnote{Here we are assuming that divergent terms should be removed by holographic renormalization. It might be worth checking this more explicitly.}. This appears reasonable since the other term in $\Pi_+$, which is proportional to $C$,
is locally related to the source $A$.

With this boundary condition on $\psi(r \to \infty)$ specified, the solution $\psi_E$ to the equations of motion which is regular in the interior is then uniquely determined. From this we can extract the corresponding $D$; we will find that $D$ and $A$ are related by a matrix $\sS$,
\be
D (k)= \sS (k) A (k)
\ee
then the boundary Euclidean two-point correlator is given by\footnote{Note that from~\eqref{vne},
\be \label{bdc}
\vev{\sO (x)}   = \int d^d y \, G_{E} (x-y) \, \ga^{\tau}\, \chi_{0} (y)
\ee
which in momentum space becomes
\be
\vev{\sO(k)}   =  G_{E } (k) \, \ga^{\tau}\, \chi_{0} (k)
\ee
The $\ga^\tau$ factor is due to that $G_E \sim \vev{\sO \sO^\da}$ rather than $\vev{\sO \bar \sO}$.
}
\be \label{roe}
G_E (k_\mu) = \sS (k_\mu) \ga^\tau \ .
\ee
Sometimes it may be easier to solve for $B$ in~\eqref{r0}, in which case introducing a matrix $\sT$
\be
B = \sT A
\ee
we then have from~\eqref{r1}
\be \label{finE}
G_E (k_\mu) = - {i \ov k^2} (2m +1)  (\ga \cdot k )  \, \sT  \, \ga^\tau \ .
\ee

To conclude this subsection, we make some further remarks:

\ben

\item Using the standard scaling argument, the identification of the source and response in~\eqref{rio} and~\eqref{ebe} implies that
the scaling dimension $\De$ of $\sO$ is related to $m$ by
\be
\De = {d \ov 2} + m
\ee
which is consistent with results obtained in~\cite{Henningson:1998cd} for pure AdS.

\item In the case of a massive scalar (as discussed in Appendix~\ref{app:M}), the $r \to \infty$ limit in the ratio ${\Pi \ov \phi_E}$ is somewhat subtle since subdominant terms in $\Pi$ and $\phi_E$ also contribute which changes the overall constant. This does not appear to happen to fermions.

\item When $0 \leq m < \ha$, all terms in $\psi_\pm$ are normalizable.  We thus can choose either  $A$ or $D$ as the source and treat the other as the corresponding response. Note that if we choose $D$ as the source term then
\be
\De = {d \ov 2} - m , \quad \to \quad {d-1 \ov 2}<  \De < {d \ov 2}
\ee
In this range the double trace operator of $\sO$ is a relevant perturbation.
The partition functions for the two alternative ways to quantize the theory should be related by a Legendre transform as one is the conjugate momentum of the other \cite{klebanov:1999}.

\item For $m=\ha$, the two terms in $\psi_-$ are degenerate. Instead one has
\be
\psi_- = r^{-{d \ov 2} - \ha} (C \log r + D)
\ee
The relations between $A,C$ in~\eqref{r1} are now replaced by
\be
C = (i \ga \cdot k ) A \ .
\ee
In this case the term proportional to $A$ is not normalizable and should be treated as the source term.

\item For even $d$, if $\chi_0$ has negative chirality, then we should be imposing boundary conditions on $\psi_-$, which in turn implies that the bulk mass term must be negative. An example illustrating this is shown in Section \ref{sec:btz} with the BTZ black hole.

\een

\subsection{Prescription for retarded correlators}

Now we simply analytically continue $\psi_E$ and~\eqref{rio}--\eqref{finE}
to Lorentzian signature, precisely as in the bosonic case with ~\eqref{dfR} and~\eqref{contGR}. We obtain the following prescription for computing retarded correlators for spinor operators:

\ben

\item  Find a solution $\psi_R(r, k_\mu)$ to the Lorentzian equations of motion which is in-falling at the horizon.

\item Expand $\psi_R (r, k_\mu)$ near $r \to \infty$ as
\be \label{r10}
\psi_{R+} = A r^{-{d \ov 2} + m} + B r^{-{d \ov 2} - m -1},
\qquad
\psi_{R-} = C r^{-{d \ov 2} + m -1} + D r^{-{d \ov 2} - m }, \quad r \to \infty
\ee
where $\psi_{R\pm}$ are spinors of definite $\Gamma^r$ eigenvalue as defined in~\eqref{deC}.

\item Then $G_R (k_\mu)$ is obtained from the analytic continuation of \eqref{roe} by
\be \label{finEx}
G_R (k_\mu) = i \sS (k_\mu) \ga^t  =  {2m +1 \ov k^2}  (\ga \cdot k )  \, \sT  \, \ga^t  \
\ee
where $\ga \cdot k = \ga^\mu k_\mu, \,k^2 = k_\mu k^\mu$ and $\sS$ and $\sT$ are defined by the relations
\be \label{rri}
D = \sS A , \qquad B = \sT A \ .
\ee

\een

We end this section by noting that causality and unitarity of the boundary CFT impose two important properties on its real-time correlators. Causality implies that the retarded correlator must be analytic in the upper half of the complex $\omega$ plane; this is easily seen to be true for correlators calculated using the prescription above. Unitarity implies that the imaginary part of the diagonal elements of the matrix $G_R$ is proportional to a spectral density and must be positive for all $\omega$; as mentioned above, the overall sign of the correlator is proportional to the normalization of the Dirac action, and one can show that a sign consistent with bulk unitarity also results in positive spectral densities in the boundary theory. 

\section{Examples} \label{sec:ex}

In this section we apply the prescription developed in last section to two exactly solvable examples, fermionic retarded correlators from pure AdS and a BTZ black hole background~\cite{btz}.

\subsection{Pure AdS} \label{sec:pureads}

The Euclidean vacuum two-point function for a spinor operator $\sO$ in a CFT, which is dual a fermonic field $\psi$ in a pure AdS, was obtained in closed form
before in~\cite{Henningson:1998cd} (see also~\cite{Mueck:1998iz,henneaux:1999}). The corresponding retarded function can then obtained from it by analytic continuation using~\eqref{contGR}. Here we calculate the retarded function directly using the prescription developed earlier as a check of our formalism.

For pure AdS, the metric is given by
\be
ds^2 = r^2 (-dt^2 + d \vec x^2)+ {dr^2 \ov r^2}, \qquad r \to + \infty \ .
\ee
with associated nonzero spin connection components given by
\be
\om_{t r} =- rdt , \quad
\om_{i r} =  r  dx^i \ .
\ee
The Dirac equation in momentum space is then given by
\be \label{Die}
r \Ga^r \p_r \psi + {i \ov r} \Ga \cdot k \psi + {d \ov 2} \Ga^r \psi
- m \psi =0
\ee
Using~\eqref{deC}, equation~\eqref{Die} now becomes coupled equations for $\psi_\pm$
\be
\psi_+ = - {i \ga \cdot k \ov k^2 }   A(-m) \psi_-, \qquad
\psi_- =   {i \ga \cdot k \ov k^2}  A(m)  \psi_+
\ee
where we have used~\eqref{evch} or~\eqref{chga} for $d$ even or odd and
\be
A (m) =r \le( r \p_r + {d \ov 2} -m \ri)
\ee
from which we obtain
\be \label{rkr}
k^2 \psi_+ =  A(-m) A(m) \psi_+ \ .
\ee
Note that~\eqref{rkr} is now a group of decoupled scalar equations, which implies that the $\sT$ matrix defined in~\eqref{rri} will be proportional to the identity matrix. Equation~\eqref{rkr} can be solved exactly using Bessel functions, and the solution satisfying the in-falling boundary condition at the horizon is
\be
\psi_{R+} = \begin{cases}  r^{-{d+1 \ov 2}} K_{m+\ha} \le({\sqrt{|\vec k|^2-\om^2 }\ov r} \ri) a_+  & k^2 > 0 \cr
                   r^{-{d+1 \ov 2}} H^{(1)}_{m+\ha} \le({\sqrt{\om^2 - |\vec k|^2} \ov r} \ri) a_+  & \om > |\vec k| \cr
                    r^{-{d+1 \ov 2}} H^{(2)}_{m+\ha} \le({\sqrt{\om^2 - |\vec k|^2} \ov r} \ri) a_+  & \om < -|\vec k|
                    \end{cases}
\ee
where $a_+$ is an arbitrary constant spinor. The story for the spacelike case $k^2 > 0$ is exactly the same as that of Euclidean correlator and the retarded correlator is real. For the timelike case $\om >  |\vec k|$, we find the corresponding $A,B$ coefficients as defined in~\eqref{r10} as
\be
A = -{1 \ov \Ga(\ha -m)} \le({k \ov 2} \ri)^{-(m+\ha)} a_+, \qquad
B =  {e^{-(m+\ha) \pi i} \ov \Ga(m+{3 \ov 2})} \le({k \ov 2} \ri)^{(m+\ha)} a_+, \quad k = \sqrt{\om^2 -|\vec k|^2}
\ee
giving
\be
\sT_{\beta}{^\al} =  \delta_{\beta}{^\al} {\Ga (-m - \ha)
\ov  \Ga (m + \ha)}  \le({k \ov 2} \ri)^{2m+1} e^{-(m+\ha) \pi i}
\ee
Using~\eqref{finEx}, we then find that
\be
G_R (k) =   {2 e^{-(m+\ha) \pi i} \ov k^2}  {\Ga (-m + \ha)
\ov  \Ga (m + \ha)}  \le({k \ov 2} \ri)^{2m+1} \le( {\ga \cdot k } \ri) \ga^t , \qquad
\om > |\vec k|
\ee
For the other possible timelike case $\om < -|\vec k|$ we simply change the phase factor $e^{-(m+\ha) \pi i}$
to $e^{(m+\ha) \pi i}$.

\subsection{BTZ Black Hole} \label{sec:btz}

We now consider fermionic correlators in a BTZ black hole background~\cite{btz}. Previous work~\cite{birmingham:2002} found that the quasinormal modes of a BTZ black hole precisely coincide with the poles in the retarded propagator of the appropriate operator of the dual theory (whose form is essentially fixed by conformal invariance). Here we will closely follow~\cite{birmingham:2002, das:1999} to solve the wave equation but slightly extend these results by finding the full correlator from the gravity side.

A BTZ black hole with a mass $M$ and angular momentum $J$ describes a boundary 2d CFT
in a sector with
\be
L_0 = \ha (M + J), \qquad \bar L_0 = \ha (M - J) \ .
\ee
The system has a finite entropy and non-vanishing left and right temperatures.
Here we are interested in probing this sector using fermionic operators with spin $s=\ha$.
Such an operator $\sO_\pm$ is characterized by conformal weights $(h_L,  h_R)$ with
\be
h_L + h_R = \De, \qquad h_L - h_R = \pm \ha \
\ee
where the $\pm$ sign denotes its chirality. As described in the previous section each $\sO_+$ (or $\sO_-$) is described by a Dirac spinor $\psi$ in the bulk, with different chiralities corresponding to different boundary conditions and opposite bulk mass. We will now find the retarded correlators of $\sO_\pm$ by solving the Dirac equation for $\psi$ in the BTZ geometry.

The metric of a BTZ black hole can be written as
\be
ds^2 = - {(r^2-r_+^2) (r^2-r_-^2) \ov r^2}  dt^2 + {r^2 dr^2 \ov (r^2-r_+^2) (r^2-r_-^2) } + r^2 \le(d \phi - {r_+r_- \ov r^2}
dt \ri)^2
\ee
where $\phi$ is an angular coordinate of period $2 \pi$. The mass, angular momentum, and
left and right moving temperature of the system are given by
\be
M = {r_+^2 + r_-^2 \ov 8 G}, \qquad  J = { r_+
r_- \ov  4 G}, \qquad T_L = \frac{r_+ - r_-}{2\pi} \qquad T_R = \frac{r_+ + r_-}{2\pi} \label{defTLR}
\ee
where $G$ is the 3d Newton constant.
%with the total temperature of the system given by
%\be
%\beta = \ha (\beta_L + \beta_R) = {2 \pi r_+ \ov r_+^2 - r_-^2}
%\ee
To solve the Dirac equation in~\eqref{btzM}, it is convenient to switch to a new coordinate system $(\rho, T, X)$
\be
r^2 = r_+^2 \cosh^2 {\rho} - r_-^2 \sinh^2 {\rho }, \qquad T + X = (r_+ + r_-)(t + \phi), \qquad T - X = (r_+ - r_-)(t-\phi)  \label{defTX}
\ee
in which the metric is
\be \label{btzM}
ds^2 = - \sinh^2 {\rho }  dT^2 + \cosh^2 {\rho }  dX^2  + d \rho^2 \
\ee
and the spin connections are given by
\be
\omega_{T\rho} = - \cosh\rho dT \qquad \omega_{X\rho} = \sinh\rho dX \ .
\ee

We will work in Fourier space on each constant-$\rho$ slice; a plane wave can be decomposed in either the $(X,T)$ or $(\phi,t)$ coordinate system:
\be
\psi  = e^{-ik_T T +i  k_X X} \psi (\rho,k_\mu)  = e^{-i\omega t + i k\phi}\psi (\rho,k_\mu)
\ee
where using \eqref{defTX} and \eqref{defTLR} we can see that the momenta in the original coordinate system $(\om,k)$ are related to $(k_T, k_X)$ by
\be
k_T + k_X = \frac{\omega + k}{2\pi T_R}, \qquad k_T - k_X = \frac{\omega - k}{2\pi T_L}
\ .
\label{relBTZft}
\ee
The Dirac equation can then be written as
\be
\left[\Gamma^{\rho}\left(\partial_{\rho} +\frac{1}{2}\left(\frac{\cosh\rho}{\sinh\rho} + \frac{\sinh\rho}{\cosh\rho}\right)\right) + i \le({k_X \Gamma^X \ov \cosh\rho} - {k_T \Gamma^{T} \ov  \sinh\rho} \ri) - m\right]\psi = 0
\ee
We choose a gamma matrix representation where $\Gamma^\rho = \sigma^3, \Gamma^T = i\sigma^2, \Ga^X = \sigma^1$ and write $\psi^T = (\psi_+, \psi_-)$. Now following \cite{das:1999} and letting
\be \label{exPs}
\psi_\pm \equiv \sqrt{\frac{\cosh\rho \pm \sinh\rho}{\cosh\rho\sinh\rho}}(\chi_1 \pm \chi_2), \qquad z =  \tanh^2\rho \ ,
\ee
then in terms of $\chi_{1,2}$ and $z$, the Dirac equation becomes
\begin{align}
2(1-z)\sqrt{z}\partial_z\chi_1 - i\left(\frac{k_T}{\sqrt{z}} + k_X\sqrt{z}\right)\chi_1 & = \left(m - \frac{1}{2}  + i(k_T + k_X)\right)\chi_2 \nonumber \\
2(1-z)\sqrt{z}\partial_z\chi_2 + i\left(\frac{k_T}{\sqrt{z}} + k_X\sqrt{z}\right)\chi_2 & = \left(m- \frac{1}{2}  - i(k_T + k_X)\right)\chi_1 \label{btzeqn} \ .
\end{align}
Note that the horizon is at $z = 0$ and the boundary at $z = 1$. It is now possible to eliminate one of the fields $\chi_{1,2}$ to obtain a second-order equation in the other field. The solutions to that equation are given in terms of hypergeometric functions. We are interested in evaluating the retarded correlator, and so we pick the solutions that are infalling at the horizon. These take the form\footnote{In order to display consistency of these solutions with the equations of motion \eqref{btzeqn}, it can be helpful to use the hypergeometric identity $-a F(a+1,b+1,c+1,z) + \frac{c}{1-z}F(a,b,c,z) + \frac{c-a}{z-1}F(a,b+1,c+1,z) = 0$ \cite{stegun}.}
\begin{align}
\chi_2 (z) & = z^{\alpha}(1-z)^{\beta}F(a,b;c;z) \nonumber \\
\chi_1 (z) & = \left(\frac{a - c}{c}\right)z^{\frac{1}{2} + \alpha}(1-z)^{\beta}F(a,b+1;c+1;z) \label{btzsoln}
\end{align}
where the parameters are
\be
\alpha  = -\frac{i k_T}{2} \qquad \beta = -\frac{1}{4} + \frac{m}{2}
\ee
and
\be
a = \ha \left(m + \ha\right) - {i \ov 2} (k_T - k_X) , \quad
b = \ha \left(m - \ha\right) - {i \ov 2} (k_T + k_X) , \quad
c = \frac{1}{2} -ik_T \ .
\ee
%
%\subsection{Extracting Correlator}

From~\eqref{exPs}, we know that at the AdS boundary $\psi$ has the following asymptotic behavior
\be
\psi_+ \sim A(1-z)^{{1 \ov 2} - {m \ov 2}} + B(1-z)^{1 + {m \ov 2}} \qquad \psi_- \sim C(1-z)^{1 - {m \ov 2}} + D(1-z)^{{1 \ov 2} + {m \ov 2}},
\ee
Our earlier analysis tells us that if $m > 0$ we can identify $A = \chi_0$ as the source and $D = \langle \sO_- \rangle$ as the response, and so the retarded correlator in this frame is given by
\be
\tilde G_R = i\frac{D}{A}
\ee
We now explicitly expand the solutions \eqref{btzsoln} near the boundary to extract the coefficients $D$, $A$. The relevant ratio works out to be
\be
\tilde G_R(k_T, k_X) =  -i\frac{\Gamma\left(\frac{1}{2} - m\right)}{\Gamma\left(\frac{1}{2}+m\right)}\frac{\Ga\left(\frac{1}{4}(1 - 2i(k_T - k_X) + 2m)\right)\Ga\left(\frac{1}{4}(3 - 2i(k_T + k_X) + 2m)\right)}{\Ga\left(\frac{1}{4}(1 - 2i(k_T + k_X) - 2m)\right)\Ga\left(\frac{1}{4}(3 - 2i(k_T - k_X) - 2m)\right)}
\ee
Using the relations \eqref{relBTZft} this can be rewritten in terms of momenta in the ($t,\phi$) coordinate system as
\be
\tilde G_R = -i\frac{\Gamma\left(\frac{1}{2} - m\right)}{\Gamma\left(\frac{1}{2}+m\right)}\frac{\Ga\left(h_L - i\frac{\omega - k}{4 \pi T_L}\right)\Ga\left(h_R - i\frac{\omega + k}{4 \pi T_R}\right)}
{\Ga\left(\tilde h_L - i\frac{\omega - k}{4\pi T_L} \right) \Ga\left(\tilde h_R - i\frac{\omega + k}{4 \pi T_R} \right)}
\label{GRtilde}
\ee
where we have introduced
\be
h_L = \frac{m}{2} + \frac{1}{4}, \qquad h_R = \frac{m}{2} + \frac{3}{4}
\ee
and
\be
\tilde h_L = -\frac{m}{2} + \frac{3}{4}, \qquad \tilde h_R = -\frac{m}{2} + \frac{1}{4} \ .
\ee
Note that~\eqref{GRtilde} has a nice factorized form for left and right sectors.
The correlator has a pole whenever either of the two gamma functions in the numerator has an argument that is a negative integer; thus we find the following two sequences of poles
\be
\omega = -k - 4\pi iT_R\left(n + h_R\right) \qquad \omega = k - 4\pi i T_L\left(n +h_L\right) \qquad n \in \mathbb{Z}_+
\ee
As demonstrated in \cite{birmingham:2002}, these two sequences of poles are precisely those appearing in the finite-temperature retarded correlator of an operator in a 2D CFT with
conformal weights $(h_L, h_R)$. Since $h_L - h_R = -\ha$, this is consistent with our expectation that $\psi$ corresponds to $\sO_-$.

If $m < 0$, we then find that $\psi_-$ is the source and $\psi_+$ is the response. Thus we are looking the correlation function of $\sO_+$.
One can immediately find the correlator from the results above, as the roles of $D$ and $A$ are reversed; we find that this results in a different pole structure consistent with the correlator of an operator of conformal dimensions $(\tilde h_L, \tilde h_R)$, indeed corresponding to a boundary spinor of positive chirality.\footnote{Note that in attempting to compare our assignment of conformal dimensions directly with \cite{birmingham:2002}, one should keep in mind two issues: $m$ in this paper is $-m$ in \cite{birmingham:2002}, and our choice of gamma matrices means that a 3d spinor $\psi_+$ with positive eigenvalue of $\Gamma^\rho$ has opposite 2d helicity here than it does in \cite{birmingham:2002}; thus our assignment of $h_L$ and $h_R$ is switched relative to them.}

Finally we note one last point which is important if one wishes to determine the overall normalization of the correlator. Note that the above expression was computed in the $(X,T)$ coordinate system; however we see from \eqref{defTX} that the relation between this and the $(t,\phi)$ coordinate system involves scaling the left-moving  coordinate by $(r_+ - r_-)$ and the right-moving coordinate by $(r_+ + r_-)$, i.e. by the left and right temperatures respectively. Thus the correlator in the $(t,\phi)$ frame is\footnote{$-1$'s in the exponents of the expression below are due to the fact that this is an expression in momentum space.}
\be
G_R(\omega,k) = (2\pi T_L)^{2h_L-1}(2\pi T_R)^{2h_R-1}\tilde{G}_R(k_T, k_X) \ .\label{GRtphi}
\ee
This rescaling is important for the extremal limit $T_L  \to 0$. In this limit,
using the Stirling formula in~\eqref{GRtilde}, we find that $\tilde G_R$ blows up
\be
\tilde G_R \to   -i\frac{\Gamma\left(\frac{1}{2} - m\right)}{\Gamma\left(\frac{1}{2}+m\right)}\frac{\Ga\left(h_R - i\frac{\omega + k}{4 \pi T_R}\right)} {\Ga\left(\tilde h_R - i\frac{\omega + k}{4 \pi T_R} \right)} \le(-i (\om -k) \ov 4 \pi T_L \ri)^{2 h_L-1},
\ee
but $G_R$ does have a finite limit
\be \label{peep}
G_R (\omega,k) =-i (2\pi T_R)^{2h_R-1} \le(-i (\om -k) \ov 2 \ri)^{2 h_L-1} \frac{\Gamma\left(\frac{1}{2} - m\right)}{\Gamma\left(\frac{1}{2}+m\right)}\frac{\Ga\left(h_R - i\frac{\omega + k}{4 \pi T_R}\right)} {\Ga\left(\tilde h_R - i\frac{\omega + k}{4 \pi T_R} \right)} \ .
\ee
This expression can be verified by direct calculation using the bulk spacetime for the extremal BTZ black hole with $T_L = 0$. Note that~\eqref{peep} again has a nice factorized form with the left-moving sector given by the vacuum expression.

\section{Conclusion} \label{sec:con}

%We have reviewed existing methods for computing realtime AdS/CFT correlators.

In this paper we showed that an intrinsic Lorentzian prescription for computing
retarded Green functions from gravity can be obtained by analytic continuation from the corresponding problem in Euclidean signature. An important message here and in our earlier work~\cite{iqbal:2008} is that, even in Lorentzian signature, for any bulk field--whether bosonic or fermionic--the corresponding conjugate momentum contains the response of the dual operator.
The field theory response can thus be expressed in terms of quantities with clear physical meaning in the bulk, making more transparent the origin of phenomena such as the universality of transport coefficients~\cite{iqbal:2008}.

We also explained in detail the issues that arise when attempting to find Lorentzian correlators of fermionic operators, and worked out simple examples to demonstrate the method.
It is easy to think of future applications of these methods. In particular, there has been relatively little investigation of fermionic correlators at finite temperature or density using gauge-gravity duality (for recent work see \cite{policastro:2008}), and it would be interesting to see whether effects such as the universality of transport coefficients have fermionic analogues. Recent work with the real-time response of spinor operators~\cite{vegh} has also found new phenomena such as the existence of a Fermi surface at finite chemical potential, and it remains to be seen whether fermionic probes can help reveal other unexplored structure in gauge-gravity duality.

%and it would be interesting to extend this analysis to other systems or to finite temperatures.

\begin{acknowledgements}
We would like to thank T.~Faulkner, V.~Kumar, J.~McGreevy, and D.~Vegh for helpful discussions and encouragement. Research supported in part by
%the Offices of Nuclear and High Energy Physics of the Office of
%Science of
the DOE
%U.S.~Department of Energy
under
contracts
%\#DE-AC02-05CH11231, \#DE-AC02-98CH10886 and
%cooperative research agreement
\#DF-FC02-94ER40818.
HL is also supported
in part by %the A.~P.~Sloan Foundation and
the U.S. Department
of Energy
OJI program.  NI is supported in part by National Science Foundation (NSF)
Graduate Fellowship 2006036498.
\end{acknowledgements}
\appendix

\section{Definitions of correlators} \label{app:A}

We start by assuming that $\sO$ is a hermitian bosonic operator; in that case our convention for the Euclidean correlator is
\be
G_E(t,\vec{x}) = \langle T_E \sO(x) \sO(0) \rangle,
\ee
where $T_E$ denotes Euclidean time ordering and at finite temperature the Euclidean time direction is taken to have period $\beta$.

Our conventions for the various realtime correlators are as follows, where $\rho$ denotes the thermal density matrix:
\begin{align}
G_R(t,\vec{x}) & = i\theta(t)\tr(\rho [\sO(t,\vec{x}),\sO(0)]) \\
G_A(t,\vec{x}) & = -i\theta(-t)\tr(\rho [\sO(t,\vec{x}),\sO(0)])
\end{align}
Note that with this sign convention the imaginary part of $G_R$ is equal to $\pi$ multiplied by the spectral density, and thus is \emph{positive} definite. Another common convention is to use the opposite sign for both $G_R$ and $G_A$; in particular, this was used in our earlier work \cite{iqbal:2008}.

For $\sO$ fermionic and complex we use the following conventions
\begin{align}
G_E(t,\vec{x}) & = \langle T_E \sO(x) \sO^{\dagger}(0) \rangle \label{euclidfermi},\\
G_R(t,\vec{x}) & = i\theta(t)\tr(\rho \{\sO(t,\vec{x}),\sO^{\dagger}(0)\}) \\
G_A(t,\vec{x}) & = -i\theta(-t)\tr(\rho \{\sO(t,\vec{x}),\sO^{\dagger}(0)\})
\end{align}

\section{Massive fields} \label{app:M}

For massive scalar fields, the boundary limit is more subtle due to various divergences.
To be specific we consider a scalar action of the form
\be
S = - \ha \int d^{d+1} x \sqrt{-g} \,  \le((\p \phi)^2 + m^2 \phi^2 \ri)
\ee
with the background metric satisfying the standard asymptotic AdS behavior~\eqref{lagr} near the boundary.

Then $\phi_R$ has the asymptotic behavior
\be \label{asum}
\phi_R (r,k_\mu) \approx A (k_\mu) \, r^{\Delta - d} + B(k_\mu) \, r^{- \Delta} , \qquad r \to \infty
\ee
where
\be \label{dimO}
\Delta = {d \ov 2} + \nu, \qquad
\qquad \nu = \sqrt{m^2 + {d^2 \ov 4}}
\ee
Note now the field theory source should now be taken to be $A$, which differs by a power of $r$ from the boundary value of $\phi$. This implies that
\be \label{newexpO}
\langle \sO(k_\mu) \rangle_A = \lim_{r\to\infty} r^{\Delta - d} \Pi(r;k_\mu)\bigr|_{\phi_R},
\ee
where as before $\Pi = - \sqrt{-g} g^{rr} \p_r \phi$ is the canonical momentum to $\phi$. Plugging in the near-boundary expansion \eqref{asum}, we see that $\Pi$ has the asymptotic behavior
\be
\Pi (r, k_\mu) \bigr|_{\phi_R} \approx - (\De - d) A (k_\mu) r^\De + \De B (k_\mu) r^{d-\De},
\ee
and thus the extra power of $r$ in \eqref{newexpO} exactly cancels the power of $r$ in front of the subleading solution $B$, so the final answer contains a finite part that is independent of $r$. Note nevertheless that some care must be taken in the limit \cite{freedman:1998}, and the retarded Green function $G_R$ should now be written as
\be
G_R (k_\mu) = \lim_{r \to \infty} r^{2 (\De -d)} {\Pi (r, k_\mu) \bigr|_{\phi_R} \ov \phi_R (r,k_\mu)} = (2\Delta - d){B (k_\mu) \ov A(k_\mu)}
\ee
where one is instructed to extract only the finite piece on the right-hand side.

\section{Boundary terms for spinors} \label{app:B}

Here we consider various technical issues related to the variation of the Dirac action. In particular, we explain why the momentum conjugate to $\psi_+$ can truly be considered the response of the operator $\langle \bar{\sO} \rangle$ dual to $\psi$; essentially this argument is due to \cite{henneaux:1999} and we review it here for clarity. Throughout this section we assume $m > 0$; a similar argument holds for $m < 0$ with an interchange of $\psi_{\pm}$.

Consider the bulk Euclidean Dirac action
\be \label{dirac2}
S_\mathrm{bulk} =  - \int d^{d+1} x \sqrt{g} \,  (\bar \psi \Ga^M \sD_M \psi  - m \bpsi \psi)
\ee
From this action the momentum $\Pi_+$ conjugate to $\psi_+$ is
\be
\Pi_+ = -\sqrt{g g^{rr}}\bar{\psi}_- \label{mompsi}
\ee
It appears that the momentum conjugate to $\bar{\psi}$ is identically zero; however we expect this momentum to contain information regarding the field theory operator $\sO$, which certainly does not vanish in general. Upon an integration by parts, however, we can transfer the radial derivative from $\psi$ to $\bpsi$; in that case we find that the momentum conjugate to $\psi_+$ is $0$, but the momentum $\bar{\Pi}_+$ conjugate to $\bpsi_+$ is now
\be
\bar{\Pi}_+ = - \sqrt{-g g^{rr}}\psi_- \label{mombpsi}
\ee
Confusingly, however, it appears that the two expressions above cannot apply simultaneously, although our expectation from the field theory side is that neither of them should vanish. Clearly boundary terms in the action are playing an important role; we now explain precisely how to fix these boundary terms and what their effects are.

The portion of the action \eqref{dirac2} containing radial derivatives can be written explicitly as
\be \label{radderiv}
S_\mathrm{bulk} \supset  - \int d^{d+1}x \sqrt{g g^{rr}} (\bar{\psi}_-\partial_r\psi_+ - \bar{\psi}_+\partial_r\psi_-)
\ee
From this we can see that if we vary this action around a solution to the equations of motion $\psi \to \psi + \delta\psi$; the variation works out to be
\be
\delta S_\mathrm{bulk} = \mbox{bulk term} - \int_{\partial \sM} d^{d}x \sqrt{g g^{rr}}(\bpsi_-\delta\psi_+ - \bpsi_+\delta\psi_-),
\ee
where the bulk term is proportional to the equations of motion and the boundary term follows from \eqref{radderiv}. Note that the on-shell action is thus a function of both $\psi_+$ and $\psi_-$; however, this is not correct. As mentioned in the text, if we impose infalling boundary conditions we are no longer free to choose $\psi_-$, and this information must be implemented in our variational principle. This is done by adding to our action the following boundary term
\be
S_{\partial} = -\int_{\partial \sM} d^dx \sqrt{g g^{rr}}\bpsi_+\psi_-
\ee
We now see that the variation of the full action is
\be
\delta S_{\mathrm{total}} = \delta (S_\mathrm{bulk} + S_{\partial}) = -\int_{\partial \sM} d^{d}x \sqrt{g g^{rr}}(\bpsi_-\delta\psi_+ + \delta\bpsi_+\psi_-)
\ee
Thus the on-shell action no longer depends on $\psi_-$; also, note that we have\footnote{Note that the first expression refers to a derivative from the right and the second to a derivative from the left. For our application to the generating function of a QFT we use the same convention.}
\be \label{defpi}
\Pi_+ \equiv \frac{\delta S_\mathrm{total}}{\delta \psi_+} = -\sqrt{g g^{rr}}\bar{\psi}_- \qquad \bar{\Pi}_+ \equiv  \frac{\delta S_\mathrm{total}}{\delta \bar{\psi}_+} =  - \sqrt{g g^{rr}}\psi_-  ,
\ee
where we are now defining the conjugate momenta $\Pi$ to be the on-shell variation of $S_{\mathrm{total}}$. Thus we see that the naive relations \eqref{mompsi} and \eqref{mombpsi} \emph{do} indeed hold simultaneously when we use the correct boundary action. It is easy to see that if we had started with a different action related to \eqref{dirac2} by only boundary terms we would have obtained a different value of $S_\partial$ but the final answer \eqref{defpi} would have been the same. (It is shown in \cite{henneaux:1999} that a symmetric splitting of the kinetic term in \eqref{dirac2} results in the boundary term used in the original work with spinors in AdS/CFT \cite{Henningson:1998cd,Mueck:1998iz}).

\end{document}